\documentclass[twocolumn,prl,superscriptaddress]{revtex4}

\usepackage{graphicx}
\usepackage{amsmath}

\begin{document}
%\title{Octave-spanning dual frequency comb spectroscopy \\
%in the molecular fingerprint region} 
\title{Dual frequency comb spectroscopy in the molecular fingerprint region} 
%\title{Molecular fingerprinting using a robust, octave-spanning \\
%10 micron frequency comb}
%\title{Molecular fingerprinting using a robust, octave-spanning long-wave infrared frequency comb}

\author{Henry Timmers}
\thanks{These authors contributed equally to the manuscript.}
\affiliation{Time and Frequency Division, National Institute of Standards and Technology, 325 Broadway, Boulder, Colorado 80305, USA}
\author{Abijith Kowligy}
\thanks{These authors contributed equally to the manuscript.}
\affiliation{Time and Frequency Division, National Institute of Standards and Technology, 325 Broadway, Boulder, Colorado 80305, USA}
\author{Alex Lind}
\thanks{These authors contributed equally to the manuscript.}
\affiliation{Time and Frequency Division, National Institute of Standards and Technology, 325 Broadway, Boulder, Colorado 80305, USA}
\affiliation{Department of Physics, University of Colorado, 2000 Colorado Ave., Boulder, Colorado 80309, USA}
\author{Flavio C. Cruz}
\affiliation{Time and Frequency Division, National Institute of Standards and Technology, 325 Broadway, Boulder, Colorado 80305, USA}
\affiliation{Instituto de Fisica Gleb Wataghin, Universidade Estadual de Campinas, Campinas, SP, 13083-859, Brazil}
\author{Nima Nader}
\affiliation{Applied Physics Division, National Institute of Standards and Technology, 325 Broadway, Boulder, Colorado 80305, USA}
\author{Myles Silfies}
\author{Thomas K. Allison}
\affiliation{Stony Brook University, Stony Brook, New York 11794, USA}
\author{Gabriel Ycas}
\affiliation{Applied Physics Division, National Institute of Standards and Technology, 325 Broadway, Boulder, Colorado 80305, USA}
\author{Peter G. Schunemann}
\affiliation{BAE Systems, P.O. Box 868 Nashua, New Hampshire 03063, USA}
\author{Scott B. Papp}
\author{Scott A. Diddams}
\email{scott.diddams@nist.gov}
\affiliation{Time and Frequency Division, National Institute of Standards and Technology, 325 Broadway, Boulder, Colorado 80305, USA}
\affiliation{Department of Physics, University of Colorado, 2000 Colorado Ave., Boulder, Colorado 80309, USA}

%\thanks{Corresponding author: henry.timmers@nist.gov}
%\affiliation[$\dagger$]{These authors contributed equally to this work}

\begin{abstract}
Spectroscopy in the molecular fingerprint spectral region (6.5-20 $\mu$m) yields critical information on material structure for physical, chemical and biological sciences. Despite decades of interest and effort, this portion of the electromagnetic spectrum remains challenging to cover with conventional laser technologies. In this report, we present a simple and robust method for generating super-octave, optical frequency combs in the fingerprint region through intra-pulse difference frequency generation in an orientation-patterned gallium phosphide crystal. We demonstrate the utility of this unique coherent light source for high-precision, dual-comb spectroscopy in methanol and ethanol vapor.  These results highlight the potential of laser frequency combs for a wide range of molecular sensing applications, from basic molecular spectroscopy to nanoscopic imaging.
\end{abstract}

\flushbottom
\maketitle

\thispagestyle{empty}

Infrared molecular spectroscopy is a powerful technique for measuring the chemical make-up and structure of matter.  In particular, the inter-atomic degrees of freedom within a molecule lead to series of discrete, vibrational states whose resonances are unique identifiers in the long-wave infrared (LWIR) fingerprint wavelength range that spans 6.5-20 $\mu$m ($\sim$1500-500 cm$^{-1}$). For the past fifty years, Fourier transform infrared spectroscopy (FTIR) \cite{Griffiths07} using thermal light sources has been a primary tool for determining molecular structure in this spectral region, imparting a wide ranging impact in the physical, chemical, biological, and medical sciences.  %, as well as biology, medicine, atmospheric science, and microscopy. 
While laser spectroscopy in this region has been pursued over a similar epoch, widely tunable LWIR laser sources remain challenging.  More recently, optical frequency comb spectroscopy \cite{Schiller02,Keilmann04,Ideguchi13,Coddington14} has been introduced as a compelling alternative to FTIR by providing a unique combination of large spectral bandwidth, high frequency precision, and rapid data acquisition that can be integrated with cavity enhancement techniques \cite{Bernhardt09,Spaun16} or long-distance propagation \cite{Rieker14} to enhance sensitivity.  

%.... enables a broadband, spectroscopic technique referred to as dual-comb spectroscopy (DCS) \cite{Coddington14}. While the technique draws parallels with FTIR, it requires no dynamic, spatial interferometer resulting in fast acquisition, high spectral resolution, and superb signal to noise. 

Significant effort has gone into the development and spectroscopic application of infrared frequency combs \cite{Schliesser12}, with techniques including difference frequency generation (DFG) \cite{Erny07,Cruz15,Lee17}, optical parametric oscillation (OPO) \cite{Adler10, Leindecker11, Maidment16}, mode-locked quantum cascade lasers \cite{Wang09,Hugi12}, super-continuum generation \cite{Hudson17}, and Kerr micro-resonator technology \cite{Yu16}.  However, to-date most frequency comb sources have been restricted to wavelengths below 6 $\mu$m or have only been able to access discrete portions of the fingerprint region with limited resolution and accuracy. In parallel, there has also been ongoing research to generate broadband multi-terahertz pulses extending up to the LWIR region through intra-pulse DFG using mode-locked oscillators based on both Ti:Sapphire \cite{Bonvalet95,Huber00,Keilmann04} and Er:fiber platforms \cite{Riek15,Riek17}.  However, reported infrared powers from such optically rectified pulses at $>10$ MHz repetition rates have been limited to the $\mu$W scale.

In this report, we introduce a solution for bright, stabilized LWIR frequency combs based on intra-pulse DFG using a few-cycle pulse derived from commonplace and robust Er:fiber laser technology. The parametric conversion occurs within a quadratic non-linear crystal employing quasi-phasematching to enhance the light conversion efficiency into the LWIR regime. In this manner, we generate spectra containing $>100$ $\mu$W of power and spanning 4-12 $\mu$m (2500-830 cm$^{-1}$), with a clear pathway to coverage across the full fingerprint region. The present super-octave bandwidth, consists of 500,000 frequency comb modes providing a spectral resolution down to 100 MHz (0.0033 cm$^{-1}$). We illustrate the capabilities of this unique spectroscopic tool by measuring mode-resolved spectra of methanol and ethanol molecules using a second comb source for readout in a dual-comb spectroscopy (DCS) configuration \cite{Coddington14}. Our results realize improvements in resolution that are a factor of 500 beyond the only previously reported frequency comb spectroscopy in the fingerprint region \cite{Keilmann04}. Additionally, the frequency axis of the spectra we recover is precisely calibrated with respect to absolute microwave standards at fractional uncertainties below $1\times 10^{-11}$. Finally, the acquisition rate of 20 ms and %absorbance sensitivity of $7\times 10^{-7}$ Hz$^{-1/2}$ per spectral element \cite{Supplement}% 
favorable signal-to-noise ratio allows us to demonstrate dynamic sampling and tracking of gas concentrations on the second time scale.  The simplicity of our source and the potential for achieving still broader spectral coverage should be enabling for a wide-range of diagnostics of chemical and biological samples in gas, liquid, or solid phase.  Extension of the laser and nonlinear techniques introduced here could facilitate applications in astronomical heterodyne spectroscopy \cite{Hale00}, pump-probe and nonlinear microscopy \cite{Kukura07,Ideguchi13}, and scanning probe techniques \cite{Pollard15,Dazzi17} for laboratory spectro-imaging at the nano-scale.

%The demonstration of such a simple scheme for generating LWIR frequency combs holds the potential for revolutionizing precision molecular spectroscopy, enabling applications including LWIR-based remote sensing \cite{Rieker14}, microscopic and nanoscopic imaging spectroscopy \cite{Ideguchi13, Pollard15}, astronomical heterodyning \cite{Hale00} to study interstellar polycyclic aromatic hydrocarbons \cite{Kaiser15}, and pump-probe spectroscopy as a means to probe conformational molecular changes during a chemical reaction \cite{Kukura07}.

%Due to the off-set free nature of DFG-based combs, the first technique presents the most convenient source for DCS. The generation of MIR combs in the mid-wavelength infrared band (MWIR, 2-5 $\mu$m) is fairly well established, involving DFG between an Er and Yb comb originating from the same oscillator \cite{Erny07, Meek13, Cruz15}.  By exploiting quasi-phasematching in a periodically poled lithium niobate (PPLN) crystal, a high DFG conversion efficiency can be achieved into the MWIR wavelength regime.  However, the generation of an MIR comb in the long-wavelength infrared band (LWIR, 5-20 $\mu$m) is still in its infancy,  primarily due to the limited number of quasi-phasematched non-linear crystals exhibiting high non-linear coefficients and broad transparencies across the pump, signal, and idler wavelength bands.  

\begin{figure}[t]
\centering\includegraphics[width=\linewidth]{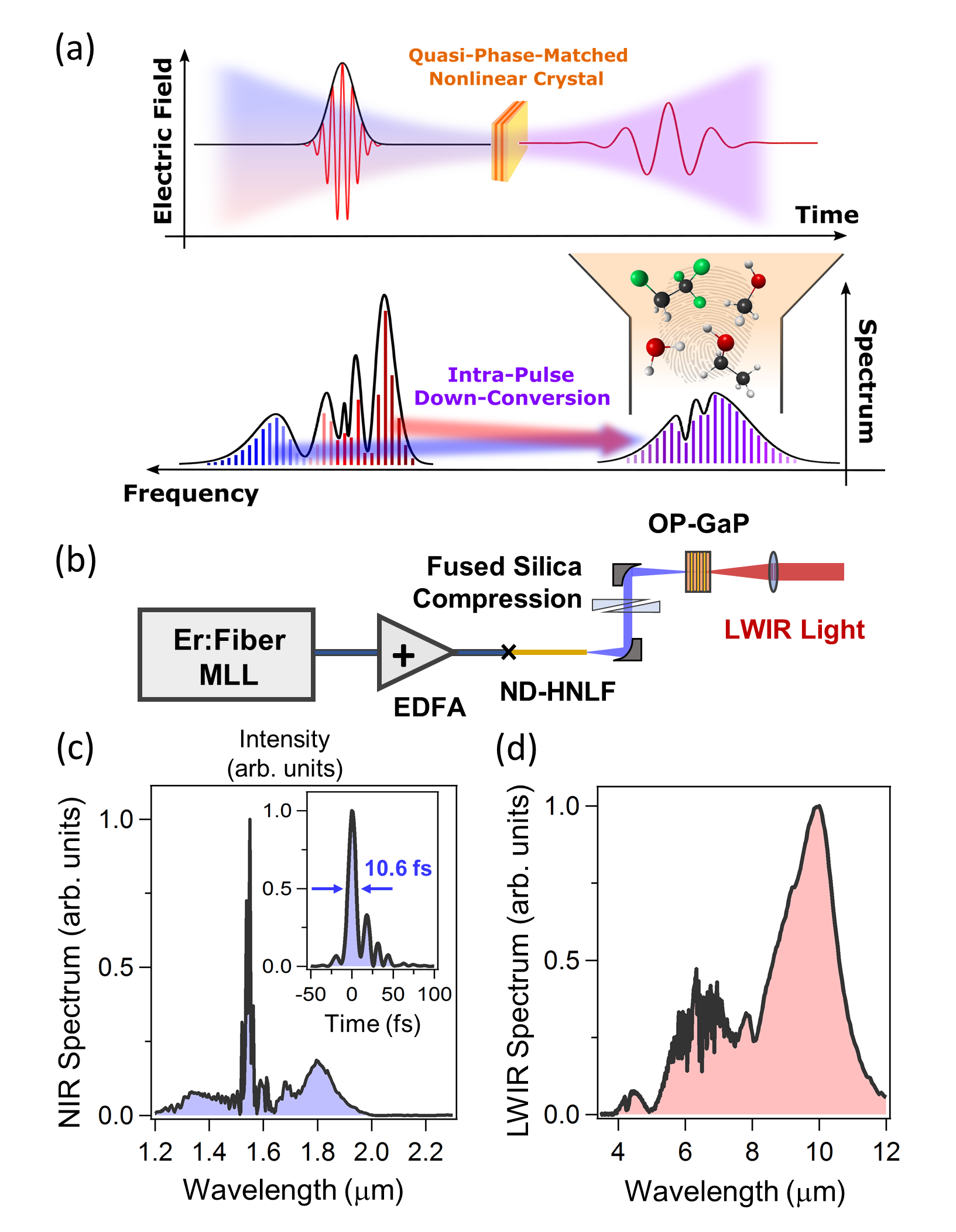}
\caption{\textbf{Experimental layout for intra-pulse DFG comb generation}.  (a) Conceptually, in the time domain (top panel) a few-cycle pulse undergoes intra-pulse DFG within a quasi-phase-matched, non-linear crystal leading to the generation of radiation with a longer optical period than the pump pulse.  In the frequency domain (bottom panel), the various comb modes within the few-cycle pulse spectrum mix within the crystal leading to the generation of down-converted light lying within the molecular fingerprint window. (b) Experimentally, the output of a Er mode-locked laser (MLL) is amplified using an Er-doped fiber amplifier (EDFA) and undergoes non-linear broadening within a ND-HNLF.  The positive chirp accumulated within the ND-HNLF is compensated using a pair of fused silica wedges, resulting in a few-cycle pump pulse. The spectrum of this few-cycle driver is shown in (c). The inset of (c) displays the measured intensity profile of the pump pulse, corresponding to a driving pulse duration of 10.6 fs.  (d) When the pump pulse is focused into an OP-GaP crystal, we generate super-octave LWIR spectra containing up to 0.25 mW of power.}
\label{Fig1}
\end{figure}

The concept and implementation of our frequency comb source is depicted in Fig. \ref{Fig1}.  %Key to the process is DFG that occurs between spectral components within the few-cycle, near-infrared pump pulse spectrum
In the time-domain [Fig. \ref{Fig1}(a), top panel], a few-cycle pulse is focused into the nonlinear crystal, resulting in a nonlinear polarization and forward emission of LWIR light having an optical period on the order of the temporal duration of the pump pulse.  In the frequency domain [Fig. \ref{Fig1}(a), lower panel], this corresponds to DFG between the spectral components within the few-cycle, near-infrared pump spectrum. %Since the process occurs in a repetitive and coherent train of few-cycle pulses, 
Since the few-cycle pulses occur as a phase-stable pulse train, the pump spectrum actually consists of a comb of frequency modes given by $\nu_{n}=f_0+n\times f_\text{rep}$, where $f_0$ is the carrier-envelope offset frequency and $f_\text{rep}$ is the repetition rate of the pulse train. The pairwise difference between the $n^{th}$ and $m^{th}$ pump modes yields a LWIR comb with frequencies, $\nu_i$, given by $\nu_{i}=(n-m)\times f_\text{rep}$. A critical and advantageous aspect of the intra-pulse difference frequency is that $f_0$ is subtracted out from the pump field, providing an offset free LWIR comb consisting of exact harmonics of $f_\text{rep}$. 
%The down-converted light lies within the molecular fingerprint regime of the LWIR spectral window (6.5-20 $\mu$m), making it an ideal tool for molecular spectroscopy. 

The experimental layout used to implement the broadband, few-cycle Er comb is shown in Fig. \ref{Fig1} (b), with complete experimental details given in the Supplement \cite{Supplement}. Briefly, femtosecond pulses generated from a commercial, 100 MHz Er:fiber mode-locked oscillator are amplified to a pulse energy of 3.5 nJ, corresponding to an average power of 350 mW.  The compressed output of the amplifier is spliced directly to a normal dispersion highly nonlinear fiber (ND-HNLF, 4 cm), where the pulse undergoes spectral broadening to generate a bandwidth of $\sim600$ nm [Fig. \ref{Fig1} (c)].  The normal dispersion of the HNLF provides the accumulation of positive chirp as the pulse propagates through the fiber that can be easily compressed using a variable length of anomalous dispersion fused silica wedges. %Finally, the broadened spectrum, containing 2.5 nJ of pulse energy, is output coupled into free space and compressed using a variable length of fused silica wedges. Since fused silica displays anomalous dispersion at wavelengths $> 1.3$ $\mu$m, it can be used to compensate for the positive chirp accumulated during ND-HNLF propagation.
The compressed pulse is characterized using frequency resolved optical gating \cite{Supplement}, and the reconstructed pulse profile [inset of Fig. \ref{Fig1} (c)] yields a temporal duration of 10.6 fs, corresponding to a 2-cycle pulse.  After compression, the pulse energy is measured to be 2.5 nJ, yielding a peak power of 250 kW. %The experimental pulse profile is well reproduced by a nonlinear Schr\"{o}dinger equation (NLSE) simulation (Fig. \ref{Fig1} (c), blue dashed curve), indicating our full understanding of this robust technique.
%, assuming broadening in the ND-HNLF fiber and compression by bulk fused silica. 

\begin{figure*}[t!]
\centering
\includegraphics[width=0.79\linewidth]{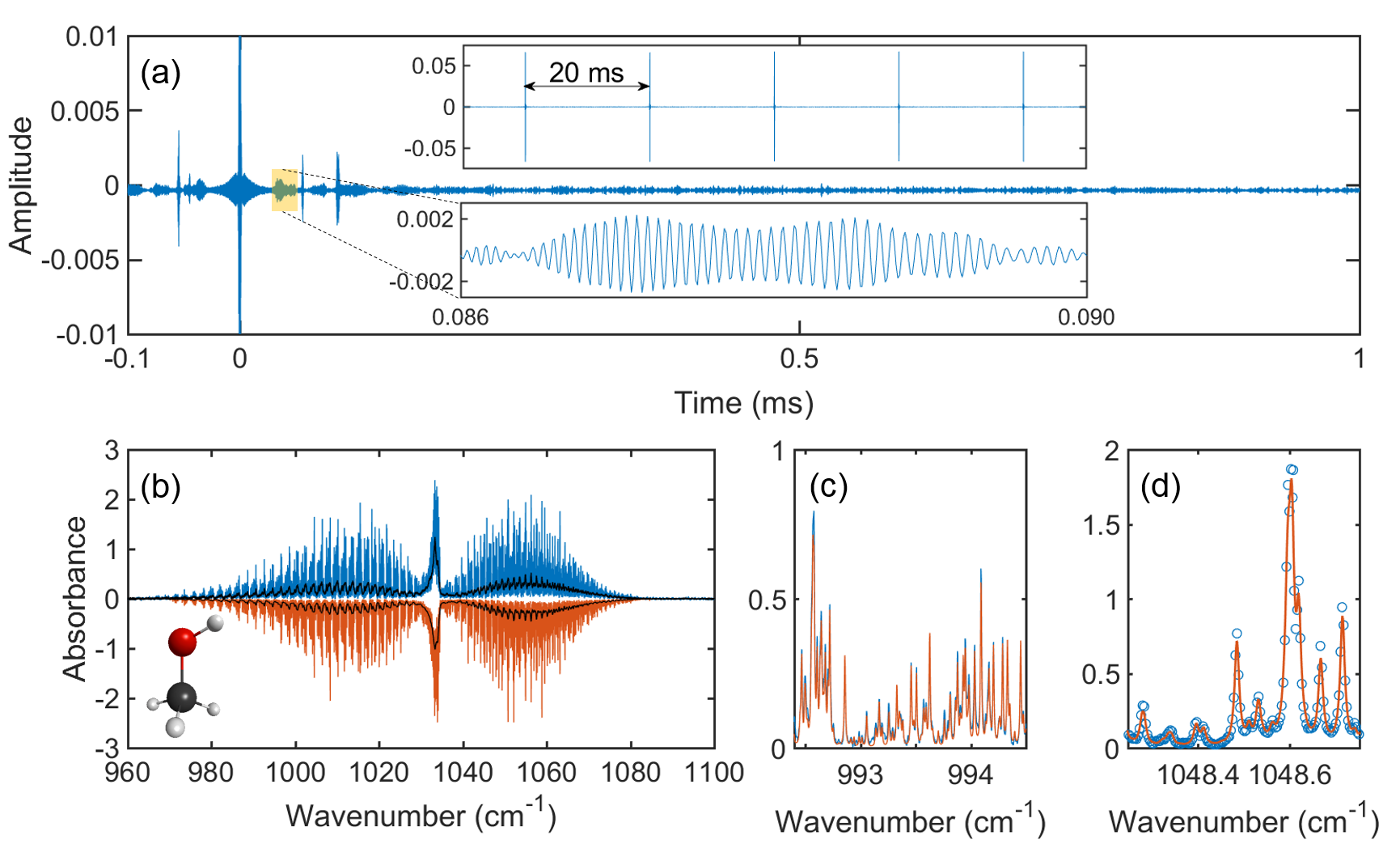}
\caption{\textbf{Dual-comb spectroscopy of methanol.} (a) In the time-domain, the multi-heterodyne signal from two LWIR frequency combs results in a periodic interferogram with a recurrence period of $1/\delta f_{\text{rep}}=$20 ms.  A sequence of 5 interferograms is shown in the inset (top) for the acquisition of comb-tooth resolved spectra. The free-induction decay signature of molecular absorption is shown in the bottom inset demonstrating the high-contrast resolution of a complete absorption oscillation cycle. (b) The measured DCS absorbance spectrum of methanol at low pressure (blue curve) compared against the methanol spectrum from the HITRAN2012 reference model \cite{HITRAN2012} (red, reflected about origin).  The overlaying black curves are the experimental and modeled spectra for methanol acquired at atmospheric pressure. (c) To demonstrate the resolution and agreement, we plot a 2 cm$^{-1}$ subset of the spectrum around 993.5 cm$^{-1}$.  (d)  The DCS spectrum arising from the phase-coherent acquisition of multiple interferograms resolves the individual 100 MHz comb modes, which are shown as open circles. } %at a full 0.0033 cm$^{-1}$ (100~MHz) resolution from a comb-tooth resolved acquisition.}
\label{Fig2}
\end{figure*}

After compression, the few-cycle pulse is focused into an orientation-patterned gallium phosphide (OP-GaP) crystal with an orientation patterning period of $\Lambda = 61.1$ $\mu$m and a thickness of 1 mm.  The OP-GaP crystal exhibits a high non-linear coefficient ($d=70.6$ pm/V) and broad transparency across the pump and generated LWIR wavelength regime, making it an ideal crystal for LWIR light conversion \cite{Lee17, Maidment16, Schunemann16}.  The generated LWIR light is collected after the OP-GaP crystal and filtered using a long-pass filter at 3.6 $\mu$m.  Using this simple set-up, we generate LWIR spectra spanning across both infrared atmospheric transmission bands (3-5 $\mu$m and 8-12 $\mu$m) and containing up to 0.25 mW of power. A typical spectrum optimized for bandwidth is shown in Fig. \ref{Fig1} (d), exhibiting over 1.5 octaves of bandwidth from 4 to 12 $\mu$m. While our measurements are limited by the 12 $\mu$m cutoff of the HgCdTe (MCT) detectors employed, additional simulations indicate that the spectrum can extend up to 25 $\mu$m \cite{Supplement}.  %However, in the current measurement and subsequent DCS, the LWIR light is limited by the spectral cut-off of the HgCdTe (MCT) detectors employed. 
%Indeed, with different nonlinear crystals, this technique can be expected to generate spectra beyond 30 $\mu$m \cite{Sell08} 

%Save this for supplement
%Since no additional branch is required for the parametric generation of the LWIR light, the comb should be intrinsically low-noise.  To confirm this, we measure the residual intensity noise (RIN) of the comb light on a HgCdTe photo-detector \cite{Supplement}.  The integrated RIN of the LWIR light is measured to be 0.08\% in the measurement window of 100 Hz to 10 MHz. In addition, we find that we can generate up to 0.25 mW of LWIR power while maintaining an super-octave bandwidth.  With such a bright source exhibiting intrinsically low intensity noise, DCS becomes a viable application for this LWIR comb light. 

To demonstrate the utility of our LWIR frequency comb, we preform high-resolution DCS of vapor phase methanol and ethanol, two commonly used solvents relevant to chemical synthesis, fuel alternatives, and commercial alcohol production.  In DCS, a second frequency comb with a slight offset in its repetition rate, $\delta f_{rep}$, is used.  The LWIR light from the two combs is combined, passes through a 15 cm molecular absorption cell, and is then photodetected.  The time-dependent interference between the two combs gives rise to a periodic interferogram with a recurrence of $1/\delta f_{rep}$, which is analogous to FTIR, but generated without the need for meter-scale mechanical delay.  
%Furthermore, a high resolution measurement of the comb spectrum can be acquired by a simple Fourier transform of the interferogram.  
%Therefore, if one or both combs pass through a molecular sample, the molecules will imprint their absorption fingerprints onto the comb spectra.  Thus, by using DCS, a high resolution reconstruction can be made of the molecular absorption profile.

The DCS interferogram measured from methanol vapor is shown in Fig. \ref{Fig2} (a).  In this experiment, we use a $\delta f_\text{rep}=50$ Hz and a temporal acquisition window of $T=1/\delta f_{rep}=20$ ms to achieve a frequency resolution of 100 MHz.  Most of the molecular information is contained in long tails and revivals following the central burst [bottom insets in Fig. \ref{Fig2} (a)].  The DCS absorbance spectrum of methanol, $A(\omega)$, is presented in Fig. \ref{Fig2} (b) and is measured as
\begin{equation}
A(\omega)=-\ln\left[\frac{|\tilde{I}_{m}(\omega)|}{|\tilde{I}_o(\omega)|}\right],
\end{equation}
where $\tilde{I}_{m}(\omega)$ and $\tilde{I}_{o}(\omega)$ are the DCS spectra with and without an absorption cell respectively and a DCS spectrum is calculated by taking a Fourier transform of the interferogram, or $\tilde{I}(\omega)=\mathcal{F}[I(t)]$.  The spectrum in Fig. \ref{Fig2} (b) corresponds to the P, Q, and R branches of the C-O stretch transition in methanol centered at 1033 cm$^{-1}$ ($\sim$9.7 $\mu$m).  We collect the methanol absorption spectrum at both atmospheric pressure (methanol partial pressure of $\sim$3 mbar, black curve), where the ro-vibrational lines are pressure broadened to $\sim$10 GHz, and at a lower background pressure of 50 mbar (methanol partial pressure of 3 mbar, blue curve), where we can resolve sub-GHz linewidths.  These collected spectra are in agreement with the methanol spectra calculated from the HITRAN2012 database \cite{HITRAN2012} (black and red curve reflected about origin, corresponding to spectra at atmospheric pressure and 50 mbar respectively).  To exemplify this agreement, we present a zoom into a subset of data at 993.5 cm$^{-1}$ in Fig. \ref{Fig2} (c).  Further, to demonstrate the low phase noise between the two LWIR combs, we additionally collect 5 interferograms in a 100 ms window, and average 5000 recurrences [upper inset of Fig. \ref{Fig2} (a)].  With multiple interferograms, we resolve the individual comb teeth, which are indicated by the circle markers in Fig. \ref{Fig2} (d).  It should be noted that the disagreement with HITRAN2012 could arise from the imprecision of the database itself, for which methanol spectra were recorded with known discrepancies \cite{Harrison12}.  Techniques such as ours could therefore be invaluable to resolving such issues.

\begin{figure}[t]
\centering
\includegraphics[width=0.85\linewidth]{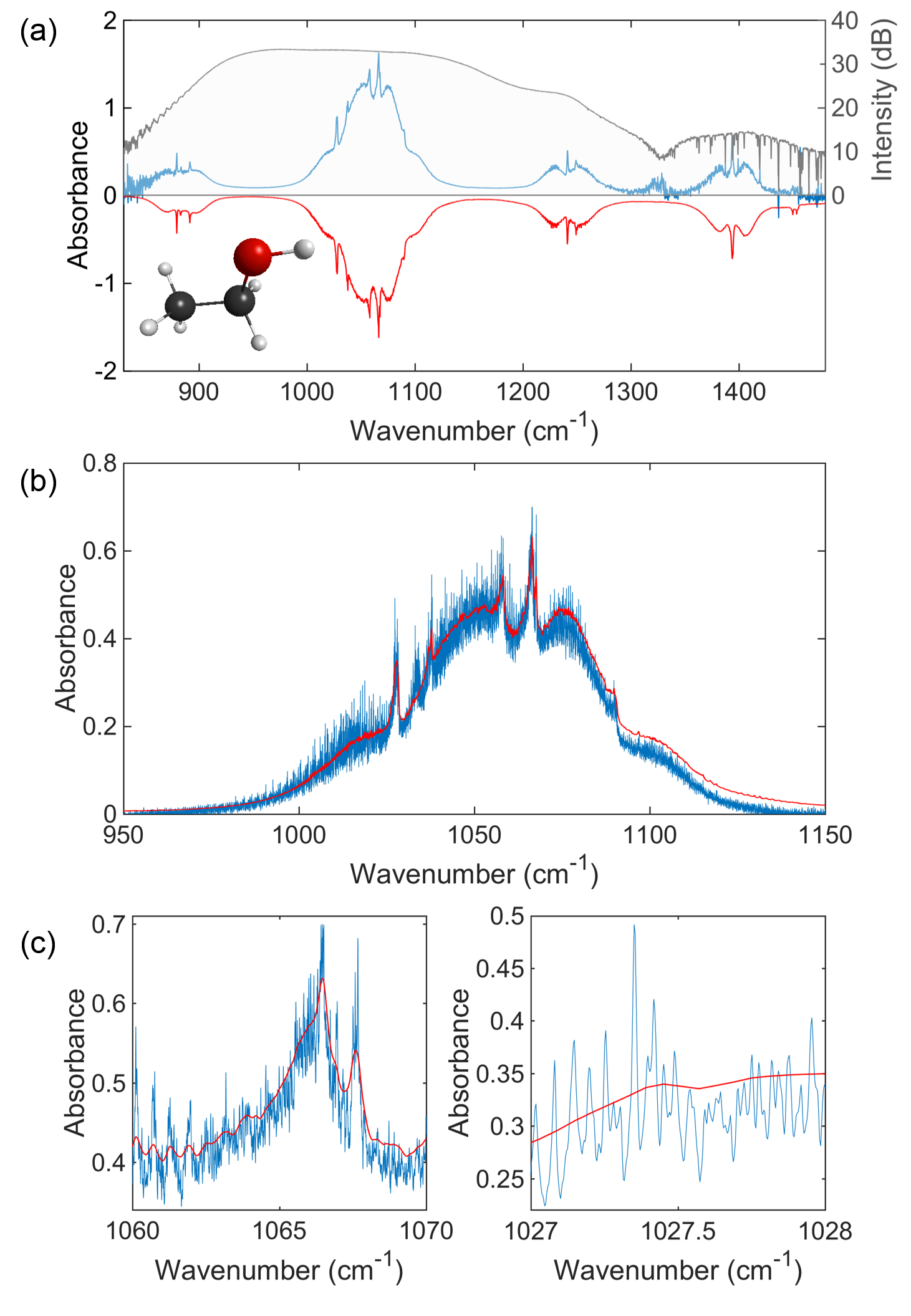}
\caption{\textbf{Dual-comb spectroscopy of ethanol.} (a) A comparison of the measured DCS spectrum from ethanol (blue curve) against the ethanol FTIR spectrum from the PNNL infrared database \cite{PNNL} (red curve, reflected about origin).  Due to the broad bandwidth of the LWIR comb, we are able to measure a DCS spectrum (black-shaded curve) spanning 700 cm$^{-1}$. (b) An ethanol DCS spectrum acquired at lower pressure with a resolution of 100 MHz (blue curve).  Using LWIR DCS, we can resolve numerous features previously unobserved by the PNNL FTIR database (red curve). (c) This level of detail is exemplified with close-ups of the LWIR spectrum at 1027.5 and 1065 cm$^{-1}$.}
\label{Fig3}
\end{figure}

While the C-O stretch transition of methanol has been largely studied using high-resolution FTIR \cite{Xu04,Harrison12}, we have found no high-resolution molecular fingerprint spectrum existing for ethanol, which exhibits four broad absorption bands across the fingerprint window.  At atmospheric pressure, we can clearly resolve these four bands over a 700 cm$^{-1}$ window, as shown in Fig. \ref{Fig3} (a) (blue spectrum).  The measured absorbance spectrum agrees well with the low-resolution FTIR spectrum from the PNNL infrared database \cite{PNNL} (red curve).  To further resolve the individual ro-vibrational lines, we lower the background pressure to $\sim$10 mbar.  This high resolution spectrum for the strongest absorption band at 1050 cm$^{-1}$ is shown in Fig. \ref{Fig3} (b) (blue curve) along with the reference PNNL spectrum (red curve).  Numerous ro-vibrational lines with sub-GHz linewidths become visible at this pressure and resolution.  Fig. \ref{Fig3} (c) presents two spectral windows that contrast the level of resolution provided by the DCS technique in comparison to previous measurements. Furthermore, by characterizing the noise on the reference spectrum [black curve in \ref{Fig3} (a)] we estimate a signal-to-noise ratio of SNR=14 Hz$^{1/2}$ for the $\mathrm{M}\sim 10^5$ modes within the central 300 cm$^{-1}$ spectral width \cite{Supplement}. This yields a quality factor of $\mathrm{M\times SNR}=1.4\times 10^6$, which is unparalleled for the fingerprint region and competitive with other DCS measurements performed in the near- and mid-infrared \cite{Coddington14}. 

\begin{figure}[t]
\centering\includegraphics[width=\linewidth]{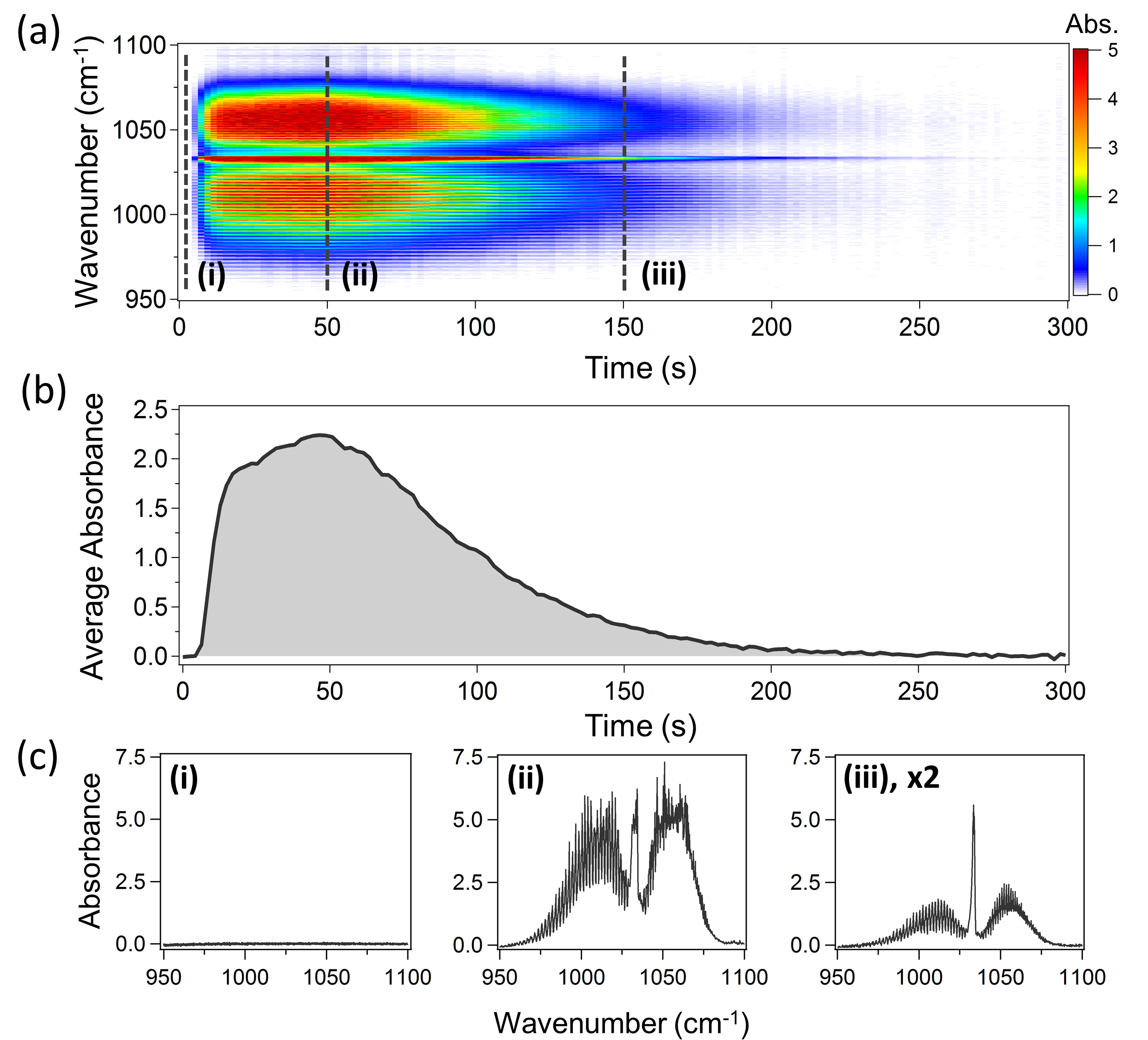}
\caption{\textbf{Time-series analysis of methanol evaporation.} (a) A time series of DCS spectra collected during the evaporation of methanol within the absorption gas cell. (b) The measured average absorbance demonstrates the ability of LWIR-based DCS in detecting transient changes in a reaction occurring on a few second timescale. (c) Spectral line outs of the time series taken along the lines denoted by \textit{(i)}, \textit{(ii)}, and \textit{(iii)} in (a).  Even with a short averaging time of 2 s per time step, we can achieve simultaneous high spectral resolution (7 GHz) and high signal-to-noise (SNR=100). }
\label{Fig4}
\end{figure}

% cite supplement in first sentence
Due to the good SNR of the LWIR combs, DCS can be performed over fairly short averaging times.  To demonstrate this, we collect a time series of DCS spectra averaged over 100 interferograms ($\sim 2$ s averaging time) in Fig. \ref{Fig4} (a).  We again use $\delta f_{rep}=50$ Hz, however, we reduce the DCS acquisition window to 285 $\mu$s, corresponding to a spectral resolution of 7 GHz.  In contrast to the static measurements of Fig. \ref{Fig2}, here we place a drop of methanol into the open gas cell and record the methanol absorption spectrum as it evaporates.  The total time series window is 350 s, during which we can capture the initial evaporation of the methanol leading to a local increase in the partial pressure of methanol as well as the slow diffusion of the vapor into the atmosphere.  These dynamics are captured in the average absorbance plot in Fig. \ref{Fig4} (b).  From this plot, we measure the time constants for the initial evaporation and the ensuing diffusion events to be $\tau_1=6.1\pm0.4$ s and $\tau_2=46.5\pm0.5$ s, respectively.  This demonstrates that LWIR DCS can be used to monitor transient changes in absorption spectra on reaction timescales corresponding to a few seconds. Fig. \ref{Fig4} (c) plots absorbance spectra at various time points depicted by the dashed lines in Fig. \ref{Fig4} (b).  Despite the fast acquisition time, we can still resolve the narrow absorption features of methanol with a signal-to-noise ratio of up to 100.

In conclusion, we present a robust scheme for generating super-octave spanning LWIR frequency combs driven by a robust, few-cycle Er-pump infrastructure.  The combination of high resolution and fast acquisition DCS demonstrates the potential of these robust, low-noise LWIR combs for a wide range of molecular fingerprint applications, both in lab-based and field-based instruments.  Going forward, we envision optically up-converting the LWIR combs via an electro-optical sampling scheme \cite{Huber00,Pupeza14,Riek15,Riek17}, to implement LWIR spectroscopy without the need for cryogenically cooled HgCdTe photo-detectors.

This research was supported in part by the DARPA SCOUT program and the NIST Greenhouse Gas and Climate Science program.  H. Timmers acknowledges support from the National Research Council. We thank J. Biegert, I. Coddington, F. Giorgetta, D. Hickstein, %A. Leitenstorfer, 
and N. Newbury for valuable comments and discussions.  This work is a contribution of the U.S. Government and is not subject to copyright in the U.S.A.

% Bibliography
\bibliography{OPGaPPaper}

\begin{thebibliography}{34}
\expandafter\ifx\csname natexlab\endcsname\relax\def\natexlab#1{#1}\fi
\expandafter\ifx\csname bibnamefont\endcsname\relax
  \def\bibnamefont#1{#1}\fi
\expandafter\ifx\csname bibfnamefont\endcsname\relax
  \def\bibfnamefont#1{#1}\fi
\expandafter\ifx\csname citenamefont\endcsname\relax
  \def\citenamefont#1{#1}\fi
\expandafter\ifx\csname url\endcsname\relax
  \def\url#1{\texttt{#1}}\fi
\expandafter\ifx\csname urlprefix\endcsname\relax\def\urlprefix{URL }\fi
\providecommand{\bibinfo}[2]{#2}
\providecommand{\eprint}[2][]{\url{#2}}

\bibitem[{\citenamefont{Griffiths and de~Haseth}(2007)}]{Griffiths07}
\bibinfo{author}{\bibfnamefont{P.}~\bibnamefont{Griffiths}} \bibnamefont{and}
  \bibinfo{author}{\bibfnamefont{J.}~\bibnamefont{de~Haseth}},
  \emph{\bibinfo{title}{Fourier Transform Infrared Spectrometry}}
  (\bibinfo{publisher}{John Wiley \& Sons, Inc.}, \bibinfo{year}{2007}),
  \bibinfo{edition}{2nd} ed.

\bibitem[{\citenamefont{Schiller}(2002)}]{Schiller02}
\bibinfo{author}{\bibfnamefont{S.}~\bibnamefont{Schiller}},
  \bibinfo{journal}{Opt. Lett.} \textbf{\bibinfo{volume}{27}},
  \bibinfo{pages}{766} (\bibinfo{year}{2002}).

\bibitem[{\citenamefont{Keilmann et~al.}(2004)\citenamefont{Keilmann, Gohle,
  and Holzwarth}}]{Keilmann04}
\bibinfo{author}{\bibfnamefont{F.}~\bibnamefont{Keilmann}},
  \bibinfo{author}{\bibfnamefont{C.}~\bibnamefont{Gohle}}, \bibnamefont{and}
  \bibinfo{author}{\bibfnamefont{R.}~\bibnamefont{Holzwarth}},
  \bibinfo{journal}{Opt. Lett.} \textbf{\bibinfo{volume}{29}},
  \bibinfo{pages}{1542} (\bibinfo{year}{2004}).

\bibitem[{\citenamefont{Ideguchi et~al.}(2013)\citenamefont{Ideguchi, Holzner,
  Bernhardt, Guelachvili, Picqu\'{e}, and H\"{a}nsch}}]{Ideguchi13}
\bibinfo{author}{\bibfnamefont{T.}~\bibnamefont{Ideguchi}},
  \bibinfo{author}{\bibfnamefont{S.}~\bibnamefont{Holzner}},
  \bibinfo{author}{\bibfnamefont{B.}~\bibnamefont{Bernhardt}},
  \bibinfo{author}{\bibfnamefont{G.}~\bibnamefont{Guelachvili}},
  \bibinfo{author}{\bibfnamefont{N.}~\bibnamefont{Picqu\'{e}}},
  \bibnamefont{and} \bibinfo{author}{\bibfnamefont{T.~W.}
  \bibnamefont{H\"{a}nsch}}, \bibinfo{journal}{Nature}
  \textbf{\bibinfo{volume}{502}}, \bibinfo{pages}{355} (\bibinfo{year}{2013}).

\bibitem[{\citenamefont{Coddington et~al.}(2016)\citenamefont{Coddington,
  Newbury, and Swann}}]{Coddington14}
\bibinfo{author}{\bibfnamefont{I.}~\bibnamefont{Coddington}},
  \bibinfo{author}{\bibfnamefont{N.}~\bibnamefont{Newbury}}, \bibnamefont{and}
  \bibinfo{author}{\bibfnamefont{W.}~\bibnamefont{Swann}},
  \bibinfo{journal}{Optica} \textbf{\bibinfo{volume}{3}}, \bibinfo{pages}{414}
  (\bibinfo{year}{2016}).

\bibitem[{\citenamefont{Bernhardt et~al.}(2009)\citenamefont{Bernhardt, Ozawa,
  Jacquet, Jacquey, Kobayashi, Udem, Holzwarth, Guelachvili, H\"{a}nsch, and
  Picqu\'{e}}}]{Bernhardt09}
\bibinfo{author}{\bibfnamefont{B.}~\bibnamefont{Bernhardt}},
  \bibinfo{author}{\bibfnamefont{A.}~\bibnamefont{Ozawa}},
  \bibinfo{author}{\bibfnamefont{P.}~\bibnamefont{Jacquet}},
  \bibinfo{author}{\bibfnamefont{M.}~\bibnamefont{Jacquey}},
  \bibinfo{author}{\bibfnamefont{Y.}~\bibnamefont{Kobayashi}},
  \bibinfo{author}{\bibfnamefont{T.}~\bibnamefont{Udem}},
  \bibinfo{author}{\bibfnamefont{R.}~\bibnamefont{Holzwarth}},
  \bibinfo{author}{\bibfnamefont{G.}~\bibnamefont{Guelachvili}},
  \bibinfo{author}{\bibfnamefont{T.~W.} \bibnamefont{H\"{a}nsch}},
  \bibnamefont{and}
  \bibinfo{author}{\bibfnamefont{N.}~\bibnamefont{Picqu\'{e}}},
  \bibinfo{journal}{Nature Photonics} \textbf{\bibinfo{volume}{4}},
  \bibinfo{pages}{55} (\bibinfo{year}{2009}).

\bibitem[{\citenamefont{Spaun et~al.}(2016)\citenamefont{Spaun, Changala,
  Patterson, Bjork, Heckl, Doyle, and Ye}}]{Spaun16}
\bibinfo{author}{\bibfnamefont{B.}~\bibnamefont{Spaun}},
  \bibinfo{author}{\bibfnamefont{P.~B.} \bibnamefont{Changala}},
  \bibinfo{author}{\bibfnamefont{D.}~\bibnamefont{Patterson}},
  \bibinfo{author}{\bibfnamefont{B.~J.} \bibnamefont{Bjork}},
  \bibinfo{author}{\bibfnamefont{O.~H.} \bibnamefont{Heckl}},
  \bibinfo{author}{\bibfnamefont{J.~M.} \bibnamefont{Doyle}}, \bibnamefont{and}
  \bibinfo{author}{\bibfnamefont{J.}~\bibnamefont{Ye}},
  \bibinfo{journal}{Nature} \textbf{\bibinfo{volume}{533}},
  \bibinfo{pages}{517–520} (\bibinfo{year}{2016}).

\bibitem[{\citenamefont{Rieker et~al.}(2014)\citenamefont{Rieker, Giorgetta,
  Swann, Kofler, Zolot, Sinclair, Baumann, Cromer, Petron, Sweeney
  et~al.}}]{Rieker14}
\bibinfo{author}{\bibfnamefont{G.~B.} \bibnamefont{Rieker}},
  \bibinfo{author}{\bibfnamefont{F.~R.} \bibnamefont{Giorgetta}},
  \bibinfo{author}{\bibfnamefont{W.~C.} \bibnamefont{Swann}},
  \bibinfo{author}{\bibfnamefont{J.}~\bibnamefont{Kofler}},
  \bibinfo{author}{\bibfnamefont{A.~M.} \bibnamefont{Zolot}},
  \bibinfo{author}{\bibfnamefont{L.~C.} \bibnamefont{Sinclair}},
  \bibinfo{author}{\bibfnamefont{E.}~\bibnamefont{Baumann}},
  \bibinfo{author}{\bibfnamefont{C.}~\bibnamefont{Cromer}},
  \bibinfo{author}{\bibfnamefont{G.}~\bibnamefont{Petron}},
  \bibinfo{author}{\bibfnamefont{C.}~\bibnamefont{Sweeney}},
  \bibnamefont{et~al.}, \bibinfo{journal}{Optica} \textbf{\bibinfo{volume}{1}},
  \bibinfo{pages}{290} (\bibinfo{year}{2014}).

\bibitem[{\citenamefont{Schliesser et~al.}(2012)\citenamefont{Schliesser,
  Picqu\'{e}, and H\"{a}nsch}}]{Schliesser12}
\bibinfo{author}{\bibfnamefont{A.}~\bibnamefont{Schliesser}},
  \bibinfo{author}{\bibfnamefont{N.}~\bibnamefont{Picqu\'{e}}},
  \bibnamefont{and} \bibinfo{author}{\bibfnamefont{T.~W.}
  \bibnamefont{H\"{a}nsch}}, \bibinfo{journal}{Nat. Photon.}
  \textbf{\bibinfo{volume}{6}}, \bibinfo{pages}{440} (\bibinfo{year}{2012}).

\bibitem[{\citenamefont{Erny et~al.}(2007)\citenamefont{Erny, Moutzouris,
  Biegert, K\"{u}hlke, Adler, Leitenstorfer, and Keller}}]{Erny07}
\bibinfo{author}{\bibfnamefont{C.}~\bibnamefont{Erny}},
  \bibinfo{author}{\bibfnamefont{K.}~\bibnamefont{Moutzouris}},
  \bibinfo{author}{\bibfnamefont{J.}~\bibnamefont{Biegert}},
  \bibinfo{author}{\bibfnamefont{D.}~\bibnamefont{K\"{u}hlke}},
  \bibinfo{author}{\bibfnamefont{F.}~\bibnamefont{Adler}},
  \bibinfo{author}{\bibfnamefont{A.}~\bibnamefont{Leitenstorfer}},
  \bibnamefont{and} \bibinfo{author}{\bibfnamefont{U.}~\bibnamefont{Keller}},
  \bibinfo{journal}{Opt. Lett.} \textbf{\bibinfo{volume}{32}},
  \bibinfo{pages}{1138} (\bibinfo{year}{2007}).

\bibitem[{\citenamefont{Cruz et~al.}(2015)\citenamefont{Cruz, Maser, Johnson,
  Ycas, Klose, Giorgetta, Coddington, and Diddams}}]{Cruz15}
\bibinfo{author}{\bibfnamefont{F.~C.} \bibnamefont{Cruz}},
  \bibinfo{author}{\bibfnamefont{D.~L.} \bibnamefont{Maser}},
  \bibinfo{author}{\bibfnamefont{T.}~\bibnamefont{Johnson}},
  \bibinfo{author}{\bibfnamefont{G.}~\bibnamefont{Ycas}},
  \bibinfo{author}{\bibfnamefont{A.}~\bibnamefont{Klose}},
  \bibinfo{author}{\bibfnamefont{F.~R.} \bibnamefont{Giorgetta}},
  \bibinfo{author}{\bibfnamefont{I.}~\bibnamefont{Coddington}},
  \bibnamefont{and} \bibinfo{author}{\bibfnamefont{S.~A.}
  \bibnamefont{Diddams}}, \bibinfo{journal}{Opt. Express}
  \textbf{\bibinfo{volume}{23}}, \bibinfo{pages}{246685}
  (\bibinfo{year}{2015}).

\bibitem[{\citenamefont{Lee et~al.}(2017)\citenamefont{Lee, Hensley,
  Schunemann, and Fermann}}]{Lee17}
\bibinfo{author}{\bibfnamefont{K.~F.} \bibnamefont{Lee}},
  \bibinfo{author}{\bibfnamefont{C.~J.} \bibnamefont{Hensley}},
  \bibinfo{author}{\bibfnamefont{P.~G.} \bibnamefont{Schunemann}},
  \bibnamefont{and} \bibinfo{author}{\bibfnamefont{M.~E.}
  \bibnamefont{Fermann}}, \bibinfo{journal}{Opt. Express}
  \textbf{\bibinfo{volume}{25}}, \bibinfo{pages}{17411} (\bibinfo{year}{2017}).

\bibitem[{\citenamefont{Adler et~al.}(2010)\citenamefont{Adler, Maslowski,
  Foltynowicz, Cossel, Briles, Hartl, and Ye}}]{Adler10}
\bibinfo{author}{\bibfnamefont{F.}~\bibnamefont{Adler}},
  \bibinfo{author}{\bibfnamefont{P.}~\bibnamefont{Maslowski}},
  \bibinfo{author}{\bibfnamefont{A.}~\bibnamefont{Foltynowicz}},
  \bibinfo{author}{\bibfnamefont{K.~C.} \bibnamefont{Cossel}},
  \bibinfo{author}{\bibfnamefont{T.~C.} \bibnamefont{Briles}},
  \bibinfo{author}{\bibfnamefont{I.}~\bibnamefont{Hartl}}, \bibnamefont{and}
  \bibinfo{author}{\bibfnamefont{J.}~\bibnamefont{Ye}}, \bibinfo{journal}{Opt.
  Express} \textbf{\bibinfo{volume}{18}}, \bibinfo{pages}{21861}
  (\bibinfo{year}{2010}).

\bibitem[{\citenamefont{Leindecker et~al.}(2011)\citenamefont{Leindecker,
  Marandi, Byer, and Vodopyanov}}]{Leindecker11}
\bibinfo{author}{\bibfnamefont{N.}~\bibnamefont{Leindecker}},
  \bibinfo{author}{\bibfnamefont{A.}~\bibnamefont{Marandi}},
  \bibinfo{author}{\bibfnamefont{R.~L.} \bibnamefont{Byer}}, \bibnamefont{and}
  \bibinfo{author}{\bibfnamefont{K.~L.} \bibnamefont{Vodopyanov}},
  \bibinfo{journal}{Opt. Ex.} \textbf{\bibinfo{volume}{19}},
  \bibinfo{pages}{6296} (\bibinfo{year}{2011}).

\bibitem[{\citenamefont{Maidment et~al.}(2016)\citenamefont{Maidment,
  Schunemann, and Reid}}]{Maidment16}
\bibinfo{author}{\bibfnamefont{L.}~\bibnamefont{Maidment}},
  \bibinfo{author}{\bibfnamefont{P.~G.} \bibnamefont{Schunemann}},
  \bibnamefont{and} \bibinfo{author}{\bibfnamefont{D.~T.} \bibnamefont{Reid}},
  \bibinfo{journal}{Opt. Lett.} \textbf{\bibinfo{volume}{41}},
  \bibinfo{pages}{4261} (\bibinfo{year}{2016}).

\bibitem[{\citenamefont{Wang et~al.}(2009)\citenamefont{Wang, Kuznetsova,
  Gkortsas, Diehl, K\"{a}rtner, Belkin, Belyanin, Li, Ham, Schneider
  et~al.}}]{Wang09}
\bibinfo{author}{\bibfnamefont{C.~Y.} \bibnamefont{Wang}},
  \bibinfo{author}{\bibfnamefont{L.}~\bibnamefont{Kuznetsova}},
  \bibinfo{author}{\bibfnamefont{V.~M.} \bibnamefont{Gkortsas}},
  \bibinfo{author}{\bibfnamefont{L.}~\bibnamefont{Diehl}},
  \bibinfo{author}{\bibfnamefont{F.~X.} \bibnamefont{K\"{a}rtner}},
  \bibinfo{author}{\bibfnamefont{M.~A.} \bibnamefont{Belkin}},
  \bibinfo{author}{\bibfnamefont{A.}~\bibnamefont{Belyanin}},
  \bibinfo{author}{\bibfnamefont{X.}~\bibnamefont{Li}},
  \bibinfo{author}{\bibfnamefont{D.}~\bibnamefont{Ham}},
  \bibinfo{author}{\bibfnamefont{H.}~\bibnamefont{Schneider}},
  \bibnamefont{et~al.}, \bibinfo{journal}{Opt. Ex.}
  \textbf{\bibinfo{volume}{17}}, \bibinfo{pages}{12929} (\bibinfo{year}{2009}).

\bibitem[{\citenamefont{Hugi et~al.}(2012)\citenamefont{Hugi, Villares, Blaser,
  Liu, and Faist}}]{Hugi12}
\bibinfo{author}{\bibfnamefont{A.}~\bibnamefont{Hugi}},
  \bibinfo{author}{\bibfnamefont{G.}~\bibnamefont{Villares}},
  \bibinfo{author}{\bibfnamefont{S.}~\bibnamefont{Blaser}},
  \bibinfo{author}{\bibfnamefont{H.~C.} \bibnamefont{Liu}}, \bibnamefont{and}
  \bibinfo{author}{\bibfnamefont{J.}~\bibnamefont{Faist}},
  \bibinfo{journal}{Nature} \textbf{\bibinfo{volume}{492}},
  \bibinfo{pages}{229} (\bibinfo{year}{2012}).

\bibitem[{\citenamefont{Hudson et~al.}(2017)\citenamefont{Hudson, Antipov, Li,
  Alamgir, Hu, Amraoui, Messaddeq, Rochette, Jackson, and Fuerbach}}]{Hudson17}
\bibinfo{author}{\bibfnamefont{D.~D.} \bibnamefont{Hudson}},
  \bibinfo{author}{\bibfnamefont{S.}~\bibnamefont{Antipov}},
  \bibinfo{author}{\bibfnamefont{L.}~\bibnamefont{Li}},
  \bibinfo{author}{\bibfnamefont{I.}~\bibnamefont{Alamgir}},
  \bibinfo{author}{\bibfnamefont{T.}~\bibnamefont{Hu}},
  \bibinfo{author}{\bibfnamefont{M.~E.} \bibnamefont{Amraoui}},
  \bibinfo{author}{\bibfnamefont{Y.}~\bibnamefont{Messaddeq}},
  \bibinfo{author}{\bibfnamefont{M.}~\bibnamefont{Rochette}},
  \bibinfo{author}{\bibfnamefont{S.~D.} \bibnamefont{Jackson}},
  \bibnamefont{and} \bibinfo{author}{\bibfnamefont{A.}~\bibnamefont{Fuerbach}},
  \bibinfo{journal}{Optica} \textbf{\bibinfo{volume}{4}}, \bibinfo{pages}{1163}
  (\bibinfo{year}{2017}).

\bibitem[{\citenamefont{Yu et~al.}(2016)\citenamefont{Yu, Okawachi, Griffith,
  Lipson, and Gaeta}}]{Yu16}
\bibinfo{author}{\bibfnamefont{M.}~\bibnamefont{Yu}},
  \bibinfo{author}{\bibfnamefont{Y.}~\bibnamefont{Okawachi}},
  \bibinfo{author}{\bibfnamefont{A.~G.} \bibnamefont{Griffith}},
  \bibinfo{author}{\bibfnamefont{M.}~\bibnamefont{Lipson}}, \bibnamefont{and}
  \bibinfo{author}{\bibfnamefont{A.~L.} \bibnamefont{Gaeta}},
  \bibinfo{journal}{Optica} \textbf{\bibinfo{volume}{3}}, \bibinfo{pages}{854}
  (\bibinfo{year}{2016}).

\bibitem[{\citenamefont{Bonvalet et~al.}(1995)\citenamefont{Bonvalet, Joffre,
  Martin, and Migus}}]{Bonvalet95}
\bibinfo{author}{\bibfnamefont{A.}~\bibnamefont{Bonvalet}},
  \bibinfo{author}{\bibfnamefont{M.}~\bibnamefont{Joffre}},
  \bibinfo{author}{\bibfnamefont{J.~L.} \bibnamefont{Martin}},
  \bibnamefont{and} \bibinfo{author}{\bibfnamefont{A.}~\bibnamefont{Migus}},
  \bibinfo{journal}{Appl. Phys. Lett.} \textbf{\bibinfo{volume}{67}},
  \bibinfo{pages}{2907} (\bibinfo{year}{1995}).

\bibitem[{\citenamefont{Huber et~al.}(2000)\citenamefont{Huber, Brodschelm,
  Tauser, and Leitenstorfer}}]{Huber00}
\bibinfo{author}{\bibfnamefont{R.}~\bibnamefont{Huber}},
  \bibinfo{author}{\bibfnamefont{A.}~\bibnamefont{Brodschelm}},
  \bibinfo{author}{\bibfnamefont{F.}~\bibnamefont{Tauser}}, \bibnamefont{and}
  \bibinfo{author}{\bibfnamefont{A.}~\bibnamefont{Leitenstorfer}},
  \bibinfo{journal}{Appl. Phys. Lett.} \textbf{\bibinfo{volume}{76}},
  \bibinfo{pages}{3191} (\bibinfo{year}{2000}).

\bibitem[{\citenamefont{Riek et~al.}(2015)\citenamefont{Riek, Seletskiy,
  Moskalenko, Schmidt, Krauspe, Eckart, Eggert, Burkard, and
  Leitenstorfer}}]{Riek15}
\bibinfo{author}{\bibfnamefont{C.}~\bibnamefont{Riek}},
  \bibinfo{author}{\bibfnamefont{D.~V.} \bibnamefont{Seletskiy}},
  \bibinfo{author}{\bibfnamefont{A.~S.} \bibnamefont{Moskalenko}},
  \bibinfo{author}{\bibfnamefont{J.~F.} \bibnamefont{Schmidt}},
  \bibinfo{author}{\bibfnamefont{P.}~\bibnamefont{Krauspe}},
  \bibinfo{author}{\bibfnamefont{S.}~\bibnamefont{Eckart}},
  \bibinfo{author}{\bibfnamefont{S.}~\bibnamefont{Eggert}},
  \bibinfo{author}{\bibfnamefont{G.}~\bibnamefont{Burkard}}, \bibnamefont{and}
  \bibinfo{author}{\bibfnamefont{A.}~\bibnamefont{Leitenstorfer}},
  \bibinfo{journal}{Science} \textbf{\bibinfo{volume}{350}},
  \bibinfo{pages}{420} (\bibinfo{year}{2015}).

\bibitem[{\citenamefont{Riek et~al.}(2017)\citenamefont{Riek, Sulzer, Seeger,
  Moskalenko, Burkard, Seletskiy, and Leitenstorfer}}]{Riek17}
\bibinfo{author}{\bibfnamefont{C.}~\bibnamefont{Riek}},
  \bibinfo{author}{\bibfnamefont{P.}~\bibnamefont{Sulzer}},
  \bibinfo{author}{\bibfnamefont{M.}~\bibnamefont{Seeger}},
  \bibinfo{author}{\bibfnamefont{A.~S.} \bibnamefont{Moskalenko}},
  \bibinfo{author}{\bibfnamefont{G.}~\bibnamefont{Burkard}},
  \bibinfo{author}{\bibfnamefont{D.~V.} \bibnamefont{Seletskiy}},
  \bibnamefont{and}
  \bibinfo{author}{\bibfnamefont{A.}~\bibnamefont{Leitenstorfer}},
  \bibinfo{journal}{Nature} \textbf{\bibinfo{volume}{541}},
  \bibinfo{pages}{376} (\bibinfo{year}{2017}).

\bibitem[{\citenamefont{Hale et~al.}(2000)\citenamefont{Hale, Bester, Danchi,
  Fitelson, Hoss, Lipman, Monnier, Tuthill, and Townes}}]{Hale00}
\bibinfo{author}{\bibfnamefont{D.~D.~S.} \bibnamefont{Hale}},
  \bibinfo{author}{\bibfnamefont{M.}~\bibnamefont{Bester}},
  \bibinfo{author}{\bibfnamefont{W.~C.} \bibnamefont{Danchi}},
  \bibinfo{author}{\bibfnamefont{W.}~\bibnamefont{Fitelson}},
  \bibinfo{author}{\bibfnamefont{S.}~\bibnamefont{Hoss}},
  \bibinfo{author}{\bibfnamefont{E.~A.} \bibnamefont{Lipman}},
  \bibinfo{author}{\bibfnamefont{J.~D.} \bibnamefont{Monnier}},
  \bibinfo{author}{\bibfnamefont{P.~G.} \bibnamefont{Tuthill}},
  \bibnamefont{and} \bibinfo{author}{\bibfnamefont{C.~H.}
  \bibnamefont{Townes}}, \bibinfo{journal}{Astrophys. J.}
  \textbf{\bibinfo{volume}{537}}, \bibinfo{pages}{998} (\bibinfo{year}{2000}).

\bibitem[{\citenamefont{Kukura et~al.}(2007)\citenamefont{Kukura, McCamant, and
  Mathies}}]{Kukura07}
\bibinfo{author}{\bibfnamefont{P.}~\bibnamefont{Kukura}},
  \bibinfo{author}{\bibfnamefont{D.~W.} \bibnamefont{McCamant}},
  \bibnamefont{and} \bibinfo{author}{\bibfnamefont{R.~A.}
  \bibnamefont{Mathies}}, \bibinfo{journal}{Annu. Rev. Phys. Chem.}
  \textbf{\bibinfo{volume}{58}}, \bibinfo{pages}{461} (\bibinfo{year}{2007}).

\bibitem[{\citenamefont{Pollard et~al.}(2015)\citenamefont{Pollard, Maia,
  Raschke, and Freitas}}]{Pollard15}
\bibinfo{author}{\bibfnamefont{B.}~\bibnamefont{Pollard}},
  \bibinfo{author}{\bibfnamefont{F.~C.~B.} \bibnamefont{Maia}},
  \bibinfo{author}{\bibfnamefont{M.~B.} \bibnamefont{Raschke}},
  \bibnamefont{and} \bibinfo{author}{\bibfnamefont{R.~O.}
  \bibnamefont{Freitas}}, \bibinfo{journal}{Nano Lett.}
  \textbf{\bibinfo{volume}{16}}, \bibinfo{pages}{55} (\bibinfo{year}{2015}).

\bibitem[{\citenamefont{Dazzi and Prater}(2017)}]{Dazzi17}
\bibinfo{author}{\bibfnamefont{A.}~\bibnamefont{Dazzi}} \bibnamefont{and}
  \bibinfo{author}{\bibfnamefont{C.~B.} \bibnamefont{Prater}},
  \bibinfo{journal}{Chem. Rev.} \textbf{\bibinfo{volume}{117}},
  \bibinfo{pages}{5146} (\bibinfo{year}{2017}).

\bibitem[{Sup()}]{Supplement}
\bibinfo{note}{Online Supplementary Material}.

\bibitem[{\citenamefont{Rothman et~al.}(2013)\citenamefont{Rothman, Gordon,
  Babikov, A.Barbe, Benner, Bernath, Birk, Bizzocchi, Boudon, Brown
  et~al.}}]{HITRAN2012}
\bibinfo{author}{\bibfnamefont{L.~S.} \bibnamefont{Rothman}},
  \bibinfo{author}{\bibfnamefont{I.~E.} \bibnamefont{Gordon}},
  \bibinfo{author}{\bibfnamefont{Y.}~\bibnamefont{Babikov}},
  \bibinfo{author}{\bibnamefont{A.Barbe}},
  \bibinfo{author}{\bibfnamefont{D.~C.} \bibnamefont{Benner}},
  \bibinfo{author}{\bibfnamefont{P.~F.} \bibnamefont{Bernath}},
  \bibinfo{author}{\bibfnamefont{M.}~\bibnamefont{Birk}},
  \bibinfo{author}{\bibfnamefont{L.}~\bibnamefont{Bizzocchi}},
  \bibinfo{author}{\bibfnamefont{V.}~\bibnamefont{Boudon}},
  \bibinfo{author}{\bibfnamefont{L.~R.} \bibnamefont{Brown}},
  \bibnamefont{et~al.}, \bibinfo{journal}{J. Quant. Spectrosc. Radiat.
  Transfer} \textbf{\bibinfo{volume}{130}}, \bibinfo{pages}{4}
  (\bibinfo{year}{2013}).

\bibitem[{\citenamefont{Schunemann et~al.}(2016)\citenamefont{Schunemann,
  Zawilski, Pomeranz, Creeden, and Budni}}]{Schunemann16}
\bibinfo{author}{\bibfnamefont{P.~G.} \bibnamefont{Schunemann}},
  \bibinfo{author}{\bibfnamefont{K.~T.} \bibnamefont{Zawilski}},
  \bibinfo{author}{\bibfnamefont{L.~A.} \bibnamefont{Pomeranz}},
  \bibinfo{author}{\bibfnamefont{D.~J.} \bibnamefont{Creeden}},
  \bibnamefont{and} \bibinfo{author}{\bibfnamefont{P.~A.} \bibnamefont{Budni}},
  \bibinfo{journal}{J. Opt. Soc. Am. B} \textbf{\bibinfo{volume}{33}},
  \bibinfo{pages}{D36} (\bibinfo{year}{2016}).

\bibitem[{\citenamefont{Harrison et~al.}(2012)\citenamefont{Harrison, Allen,
  and Bernath}}]{Harrison12}
\bibinfo{author}{\bibfnamefont{J.~J.} \bibnamefont{Harrison}},
  \bibinfo{author}{\bibfnamefont{N.~D.~C.} \bibnamefont{Allen}},
  \bibnamefont{and} \bibinfo{author}{\bibfnamefont{P.~F.}
  \bibnamefont{Bernath}}, \bibinfo{journal}{J. Quant. Spectrosc. Radiat.
  Transfer} \textbf{\bibinfo{volume}{113}}, \bibinfo{pages}{2189}
  (\bibinfo{year}{2012}).

\bibitem[{\citenamefont{Sharpe et~al.}(2004)\citenamefont{Sharpe, Johnson,
  Sams, Chu, Rhoderick, and Johnson}}]{PNNL}
\bibinfo{author}{\bibfnamefont{S.~W.} \bibnamefont{Sharpe}},
  \bibinfo{author}{\bibfnamefont{T.~J.} \bibnamefont{Johnson}},
  \bibinfo{author}{\bibfnamefont{R.~L.} \bibnamefont{Sams}},
  \bibinfo{author}{\bibfnamefont{P.~M.} \bibnamefont{Chu}},
  \bibinfo{author}{\bibfnamefont{G.~C.} \bibnamefont{Rhoderick}},
  \bibnamefont{and} \bibinfo{author}{\bibfnamefont{P.~A.}
  \bibnamefont{Johnson}}, \bibinfo{journal}{Appl. Spectrosc.}
  \textbf{\bibinfo{volume}{58}}, \bibinfo{pages}{1452} (\bibinfo{year}{2004}).

\bibitem[{\citenamefont{Xu et~al.}(2004)\citenamefont{Xu, Lees, Wang, Brown,
  Kleiner, and Johns}}]{Xu04}
\bibinfo{author}{\bibfnamefont{L.-H.} \bibnamefont{Xu}},
  \bibinfo{author}{\bibfnamefont{R.~M.} \bibnamefont{Lees}},
  \bibinfo{author}{\bibfnamefont{P.}~\bibnamefont{Wang}},
  \bibinfo{author}{\bibfnamefont{L.~R.} \bibnamefont{Brown}},
  \bibinfo{author}{\bibfnamefont{I.}~\bibnamefont{Kleiner}}, \bibnamefont{and}
  \bibinfo{author}{\bibfnamefont{J.~W.~C.} \bibnamefont{Johns}},
  \bibinfo{journal}{J. Mol. Spectrosc.} \textbf{\bibinfo{volume}{228}},
  \bibinfo{pages}{453} (\bibinfo{year}{2004}).

\bibitem[{\citenamefont{Pupeza et~al.}(2015)\citenamefont{Pupeza, S\'{a}nchez,
  Zhang, Lilienfein, Seidel, Karpowicz, Paasch-Colberg, Znakovskaya, Pescher,
  Schweinberger et~al.}}]{Pupeza14}
\bibinfo{author}{\bibfnamefont{I.}~\bibnamefont{Pupeza}},
  \bibinfo{author}{\bibfnamefont{D.}~\bibnamefont{S\'{a}nchez}},
  \bibinfo{author}{\bibfnamefont{J.}~\bibnamefont{Zhang}},
  \bibinfo{author}{\bibfnamefont{N.}~\bibnamefont{Lilienfein}},
  \bibinfo{author}{\bibfnamefont{M.}~\bibnamefont{Seidel}},
  \bibinfo{author}{\bibfnamefont{N.}~\bibnamefont{Karpowicz}},
  \bibinfo{author}{\bibfnamefont{T.}~\bibnamefont{Paasch-Colberg}},
  \bibinfo{author}{\bibfnamefont{I.}~\bibnamefont{Znakovskaya}},
  \bibinfo{author}{\bibfnamefont{M.}~\bibnamefont{Pescher}},
  \bibinfo{author}{\bibfnamefont{W.}~\bibnamefont{Schweinberger}},
  \bibnamefont{et~al.}, \bibinfo{journal}{Nature Photonics}
  \textbf{\bibinfo{volume}{9}}, \bibinfo{pages}{721–724}
  (\bibinfo{year}{2015}).

\end{thebibliography}

\end{document}